\documentclass[superscriptaddress,onecolumn,secnumarabic,nobibnotes,aps,prd,showpacs,nofootinbib,12pt]{revtex4}
\usepackage{graphicx}
\usepackage{epsf}
\usepackage{bm}
\usepackage{amsmath}
\usepackage{amsfonts}
\usepackage{amssymb}
\usepackage{color}
\usepackage{subfig}

\providecommand{\U}[1]{\protect\rule{.1in}{.1in}}

\newcommand{\be}{\begin{equation}}
\newcommand{\ee}{\end{equation}}

\newcommand{\mincir}{\raise
-3.truept\hbox{\rlap{\hbox{$\sim$}}\raise4.truept\hbox{$<$}\ }}
\newcommand{\magcir}{\raise
-3.truept\hbox{\rlap{\hbox{$\sim$}}\raise4.truept\hbox{$>$}\ }}

\begin{document}

\title{Avoiding Big Rip Singularities in Phantom Scalar Field theory with
Gauss-Bonnet term}
\author{Giannis Papagiannopoulos}
\affiliation{Department of Physics, National \& Kapodistrian University of Athens,
Zografou Campus GR 157 73, Athens, Greece}
\author{Genly Leon}
\affiliation{Departamento de Matem\'{a}ticas, Universidad Cat\'{o}lica del Norte, Avda.
Angamos 0610, Casilla 1280 Antofagasta, Chile}
\affiliation{Institute of Systems Science \& Department of Mathematics, Faculty of Applied Sciences, Durban University of Technology, Durban 4000, South Africa}
\author{Andronikos Paliathanasis}
\email{anpaliat@phys.uoa.gr}
\affiliation{Institute of Systems Science \& Department of Mathematics, Faculty of Applied Sciences, Durban University of Technology, Durban 4000, South Africa}
\affiliation{Centre for Space Research, North-West University, Potchefstroom 2520, South
Africa}
\affiliation{Departamento de Matem\'{a}ticas, Universidad Cat\'{o}lica del Norte, Avda.
Angamos 0610, Casilla 1280 Antofagasta, Chile}
\affiliation{School of Technology, Woxsen University, Hyderabad 502345, Telangana, India}

\begin{abstract}
We consider a phantom scalar field coupled to the Gauss-Bonnet scalar within
a spatially flat FLRW geometry. Moreover, we assume a nonzero interaction
between the scalar field and the matter term. We perform a detailed phase space analysis using two sets of dimensionless variables.
Specifically, we introduce dimensionless variables based on the Hubble
normalization approach and a new set based on the
matter-scalar field normalization. These two sets of variables
allow for a comprehensive phase space analysis. This model supports
inflationary solutions without the Big Rip or Big Crunch singularities
appearing as asymptotic solutions. This outcome is attributed to the
presence of the Gauss-Bonnet scalar. The result remains valid even in the
absence of the interaction term.
\end{abstract}

\maketitle

\section{Introduction}

The introduction of the Gauss-Bonnet scalar within the gravitational action
integral leads to a gravitational theory known as Einstein-Gauss-Bonnet
theory \cite{pan1,rr1}, which belongs to the family of Lovelock's theories
of gravity \cite{lov1}. Lovelock's theory is a natural extension of General
Relativity in higher-dimensional manifolds. It results in second-order
gravitational field equations free from Ostrogradsky instabilities \cite%
{ost1}. Lovelock's theory reduces to General Relativity in four-dimensional
spacetime.

The Gauss-Bonnet scalar is a topological
invariant in a four-dimensional manifold, contributing a boundary term to the gravitational
action integral. Consequently, the variation principle in
Einstein-Gauss-Bonnet theory yields the same field equations as in General Relativity. However, in five- or higher-dimensional manifolds, the Gauss-Bonnet scalar introduces new geometrodynamical terms in the field equations, leading to novel phenomena, as discussed in \cite%
{gb1,gb2,gb3,gb4,gb5,gb6,gb7,gb8} and references therein.

To incorporate the effects of the Gauss-Bonnet scalar in a four-dimensional
manifold, a scalar field with a nonzero coupling function can be introduced
to the Gauss-Bonnet scalar field \cite{gbs1,gbs2,gbs3}. In this case, the
When multiplied by the coupling function, the Gauss-Bonnet term ceases to be a topological invariant. This introduces extra geometrodynamical degrees of freedom into the field equations while preserving the second-order nature of gravitational theory. This model has been extensively studied in the context of cosmic inflation \cite{gbb1,gbb2,gbb3,gbb4,gbb5}. In this framework,
cosmic acceleration is attributed to the Gauss-Bonnet scalar \cite{gbb1}.
Other cosmological applications are presented in \cite%
{kant3,kant4,in4,dn3,dn4,dn6,dn7,dn8}.

In the case of a Quintessence scalar field, the presence of the Gauss-Bonnet
term affects the dynamics so that the equation-of-state parameter can
cross the phantom divide line without resulting in future Big Rip
singularities \cite{gbs1}. In this work, we consider an
Einstein-Gauss-Bonnet scalar field model with a scalar field possessing
negative kinetic energy, i.e., a phantom field \cite{q14,qq1,qq2,qq3}.
Furthermore, we assume a coupling between the scalar field and the matter
source to enable a chameleon mechanism \cite{ch1,ch2}. Due to this
interaction, the mass of the scalar field depends on the energy density of the
matter source. The interaction between the phantom field and dark matter
allows the equation of state parameter to cross the phantom divide line
without leading to future attractors that describe Big Rip singularities
\cite{q14a,q14b}. This behavior does not occur for a true phantom field in
the absence of the interaction term \cite{q15}. In the following, we
investigate the effects of the Gauss-Bonnet term on the evolution of
physical parameters when the scalar field is modeled as a phantom field. We
generalize the analysis presented in \cite{gbs1} and extend the recent study
\cite{anph}, where the chameleon term was introduced in the
Einstein-Gauss-Bonnet theory. The structure of the article is as follows.

In Section \ref{sec2}, we introduce the gravitational model under
consideration, which is Einstein-Gauss-Bonnet scalar field cosmology.
We consider the scalar field a phantom field with negative kinetic
energy. The scalar field is also coupled to the Gauss-Bonnet scalar
and the matter source, enabling the chameleon mechanism. The gravitational
model depends on three free functions: the scalar field potential $V\left(
\phi \right) $, the scalar field coupling to the Gauss-Bonnet scalar $%
f\left( \phi \right) $ and the coupling function $g\left( \phi \right) $
which defines the interaction with the matter source. In Section \ref{sec3}, we
consider the spatially flat Friedmann--Lema\^{\i}tre--Robertson--Walker
geometry, and we derive the cosmological field equations. The evolution of
the physical variables is expressed in terms of a system of nonlinear
second-order ordinary differential equations.

The asymptotic analysis is presented in Sections \ref{sec4} and \ref{sec5}.
Specifically, for the three functions in the gravitational model, we assume $%
f\left( \phi \right) $ to be linear function, $f\left( \phi \right)
=f_{0}\phi $, $g\left( \phi \right) $ to be exponential function $g\left(
\phi \right) =e^{2\beta \phi },$ as described by the chameleon mechanism, and
for the scalar field potential, we consider the exponential, i.e., $V\left(
\phi \right) =V_{0}e^{\lambda \phi }$. For this gravitational model, in
Section \ref{sec4} introduces a new set of dimensionless variables based
on the Hubble normalization approach. The field equations are reformulated
as nonlinear first-order algebraic differential equations. Each
stationary point of this system represents an asymptotic solution of the
original system \cite{dn22,dn33,dn44,dn55}. By analyzing the stability of
these stationary points, we reconstruct the cosmological history and
evolution.

Due to the system's non-linearity, the Hubble normalization does not yield a
set of compactified variables. Therefore, in Section \ref{sec5}, we
introduce a new set of dimensionless variables within the matter-scalar
field normalization framework. Using these variables, we determine the
phase-space behavior in the infinite regime. We find that, due to the
Gauss-Bonnet scalar, no stationary points correspond to Big Rip or Big
Crunch singularities, regardless of the presence of the chameleon term.
Finally, we present our conclusions in Section \ref{sec6}.

\section{Gravitational theory}

\label{sec2}

The gravitational theory of our consideration is that of
the Einstein-Gauss-Bonnet scalar field theory described by the action integral
\cite{gbb1}%
\begin{equation}
S=\int d^{4}x\sqrt{-g}\left( R-f\left( \phi \right) G+\frac{1}{2}g^{\mu
\kappa }\nabla _{\mu }\phi \nabla _{\kappa }\phi -V\left( \phi \right)
-g\left( \phi \right) L_{m}\left( x^{\nu }\right) \right) .  \label{ai.01}
\end{equation}%
$R$ is the Ricci scalar for the four-dimensional $g_{\mu \nu }$, $G$ is the
Gauss-Bonnet scalar~defined as \cite{rr1}%
\begin{equation}
G=R^{2}-4R_{\mu \nu }R^{\mu \nu }+R_{\mu \nu \kappa \lambda }R^{\mu \nu
\kappa \lambda }.  \label{ai.02}
\end{equation}%
which is a topological invariant since $g_{\mu \nu }$ has dimension four.

Function $L_{m}\left( x^{\nu }\right) $ represents the Lagrangian function
for the matter source, and $\phi $ is a scalar field coupled to the
Gauss-Bonnet scalar $G$ as well as to the matter source $L_{m}\left( x^{\nu
}\right) $. The function $g\left( \phi \right) $ facilitates energy transfer
between the scalar field $\phi $ and the matter source $L_{m}$.

$f\left( \phi \right) $ also characterizes the coupling between the
scalar field and the Gauss-Bonnet scalar. When $f\left( \phi \right) $ is
constant, the gravitational model (\ref{ai.01}) reduces to General
Relativity. The function $f\left( \phi \right) $ is critical for ensuring a
contribution from the Gauss-Bonnet term in the gravitational model.

For the scalar field $\phi $, it is assumed to be a phantom field, meaning
it can have a negative kinetic term. This implies that it may violate the
weak energy condition, allowing the energy density to become negative. The
gravitational field equations are \cite{gbb1}%
\begin{equation}
R_{\mu \nu }-\frac{1}{2}Rg_{\mu \nu }=T_{\mu \nu }^{G}+T_{\mu \nu }^{\phi
}+f\left( \phi \right) T_{\mu \nu }^{m},  \label{ai.03}
\end{equation}%
where $T_{\mu \nu }^{m}$ is the energy-momentum tensor for the matter
source, $T_{\mu \nu }^{m}=\frac{\delta L_{m}}{\delta g_{\mu \nu }}$,$~T_{\mu
\nu }^{\phi }$ is the energy-momentum tensor for the phantom scalar field 
\begin{equation}
T_{\mu \nu }^{\phi }=-\nabla _{\mu }\phi \nabla _{\nu }\phi -g^{\mu \nu
}\left( \frac{1}{2}g^{\mu \kappa }\nabla _{\mu }\phi \nabla _{\kappa }\phi
+V\left( \phi \right) \right) ,
\end{equation}%
and $T_{\mu \nu }^{G}$ is the effective energy-momentum tensor which
attributes the geometrodynamical degrees of freedom given by the
Gauss-Bonnet scalar that is \cite{gbb1} 
\begin{align}
T_{\mu \nu }^{G}& =-4\left( \nabla _{\mu }\nabla _{\nu }f\left( \phi \right)
\right) R+8\left( \nabla _{\mu }\nabla _{\rho }f\left( \phi \right) \right)
R_{\nu }^{\rho }+8\left( \nabla _{\nu }\nabla _{\rho }f\left( \phi \right)
\right) R_{\mu }^{\rho }  \notag \\
& -8\left( g^{\kappa \rho }\nabla _{\kappa }\nabla _{\rho }f\left( \phi
\right) \right) \left( 4R_{\mu \nu }-2Rg_{\mu \nu }\right) -8\left( \nabla
_{\kappa }\nabla _{\rho }f\left( \phi \right) \right) \left( R^{\rho \kappa
}g_{\mu \nu }-R_{\mu ~~\nu }^{~\ \rho ~~\sigma }\right) ,  \label{ai.04}
\end{align}%
We observe that when $f\left( \phi \right) $ is a constant, $T_{\mu \nu
}^{G}=0$, and when $f\left( \phi \right) $ is a linear function $f=f_{0}\phi 
$, it follows%
\begin{align}
T_{\mu \nu }^{G}\left( f\left( \phi \right) \rightarrow f_{0}\phi \right) &
=-4\left( \nabla _{\mu }\nabla _{\nu }\phi \right) R+8\left( \nabla _{\mu
}\nabla _{\rho }\phi \right) R_{\nu }^{\rho }+8\left( \nabla _{\nu }\nabla
_{\rho }\phi \right) R_{\mu }^{\rho }  \notag \\
& -8\left( g^{\kappa \rho }\nabla _{\kappa }\nabla _{\rho }\phi \right)
\left( 4R_{\mu \nu }-2Rg_{\mu \nu }\right) -8\left( \nabla _{\kappa }\nabla
_{\rho }\phi \right) \left( R^{\rho \kappa }g_{\mu \nu }-R_{\mu ~~\nu }^{~\
\rho ~~\sigma }\right) .  \label{ai.05}
\end{align}

The Bianchi identity leads to the conservation equation \cite{gbb1} 
\begin{equation}
g^{\mu \nu }\nabla _{\mu }\nabla _{\nu }\phi +V\left( \phi \right) +f_{,\phi
}G+\nabla _{\mu }\left( f\left( \phi \right) \rho _{m}\right) u^{\mu
}+g\left( \phi \right) \left( \rho _{m}+p_{m}\right) \nabla _{\mu }u^{\mu
}=0,  \label{ai.06}
\end{equation}%
Without loss of generality, the latter equation can be read as follows \cite%
{gbb1} 
\begin{equation}
g^{\mu \nu }\nabla _{\mu }\nabla _{\nu }\phi +V_{,\phi }+g_{,\phi }G+2\rho
_{m}\nabla _{\mu }f\left( \phi \right) u^{\mu }=0,  \label{ai.07}
\end{equation}%
\begin{equation}
\nabla _{\mu }\left( \rho _{m}\right) u^{\mu }+\left( \rho _{m}+p_{m}\right)
\nabla _{\mu }u^{\mu }-\rho _{m}\nabla _{\mu }\ln \left( f\left( \phi
\right) \right) u^{\mu }=0.  \label{ai.08}
\end{equation}%
$\rho _{m}\nabla _{\mu }\ln \left( f\left( \phi \right) \right) u^{\mu }$
defines the interaction which gives the energy transfer between the scalar
field and matter. This term introduces the Chameleon mechanism for the scalar field.

Equation (\ref{ai.07}) is the modified Klein-Gordon equation for the phantom
scalar field. We observe that the mass of the scalar field depends on the
potential function $V\left( \phi \right) $, the energy density $\rho _{m}$
of the matter, and the Gauss-Bonnet scalar $G$. The theory depends on three
unknown functions: the scalar field potential $V\left( \phi \right) $, the
Gauss-Bonnet coupling function $f\left( \phi \right) $, and the Chameleon
mechanism $g\left( \phi \right) $.

To assess the cosmological viability of the above model and investigate its
potential to unify the early-time and late-time acceleration phases of the
universe, we performed a detailed phase-space analysis of the field
equations for an isotropic and spatially flat FLRW background geometry.

\section{FLRW\ Universe}

\label{sec3}

On very large scales, the Universe is considered to be described by the
isotropic and homogeneous spatially flat FLRW geometry with line element%
\begin{equation}
ds^{2}=-N^{2}\left( t\right) dt^{2} +a^{2}\left( t\right) \left(
dx^{2}+dy^{2}+dz^{2}\right) ,  \label{ai.09}
\end{equation}%
where $a\left( t\right) $ is the scale factor and describes the radius of
the universe, $N\left( t\right) $ is a lapse function. For the commoving
observer $u^{\mu }=\delta _{t}^{\mu }$, the expansion rate is calculated $\theta =3H$, where $H=\frac{1}{N}\frac{\dot{a}}{a}$ is the Hubble function,
and $\dot{a}=\frac{da}{dt}$.

For the matter source, we consider it to be described by an isotropic,
pressure-less fluid source, which corresponds to the dark matter of the
Universe. In this case, the Lagrangian function is given by $L_{m}=\rho
_{m0}a^{-3}$.

Thus, from the gravitational Action Integral (\ref{ai.01}) For the latter line element, we define the point-like Lagrangian \cite{dn4} 
\begin{equation}
L\left( N,a,\dot{a},\phi ,\dot{\phi}\right) =-\frac{1}{N}\left( 3a\dot{a}%
^{2}+\frac{1}{2}a^{3}\dot{\phi}^{2}\right) -N\left( a^{3}\,V(\phi )+g\left(
\phi \right) \rho _{m0}\,\right) +\frac{8\,}{N^{3}}f_{,\phi }\dot{a}^{3}\dot{%
\phi},  \label{ai.10}
\end{equation}%
The cosmological field equations are derived from the variation of the
Lagrangian with respect to the dynamical variables $N$, $a$, and $\phi$.

The variation with respect to the scale factor $a$ and the scalar field $%
\phi $ results in second-order ordinary differential equations, while the
variation with respect to the lapse function $N$ provides the constraint
equation.

Without loss of generality, we assume $N\left( t\right) =1$, and the
cosmological field equations take the form \cite{dn4}%
\begin{eqnarray}
0 &=&3H^{2}+\frac{1}{2}\dot{\phi}^{2}-\,V(\phi )+g\left( \phi \right) \rho
_{m0}a^{-3}\,-24\,f_{,\phi }H\dot{\phi},  \label{ai.11} \\
0 &=&2\dot{H}+3H^{2}-\frac{1}{2}\dot{\phi}^{2}-V\left( \phi \right)
-16\left( H^{2}+\dot{H}\right) Hf_{,\phi }\dot{\phi}-8H\left( \dot{\phi}%
^{2}f_{,\phi \phi }+f_{,\phi }\ddot{\phi}\right) ,  \label{ai.12} \\
0 &=&\ddot{\phi}+3H\dot{\phi}-V_{,\phi }-24Hf_{,\phi }\left( H^{2}+\dot{H}%
\right) -g_{,\phi }\rho _{m0}a^{-3}.  \label{ai.13}
\end{eqnarray}

The field equations can be written in the equivalent form
\begin{eqnarray}
3H^{2} &=&\rho _{eff}, \\
2\dot{H}+3H^{2} &=&-p_{eff},
\end{eqnarray}%
in which 
\begin{equation}
\rho _{eff}=-\frac{1}{2}\dot{\phi}^{2}+V(\phi )-g\left( \phi \right) \rho
_{m0}a^{-3}\,+24\,f_{,\phi }H\dot{\phi},
\end{equation}%
and%
\begin{equation}
p_{eff}=\frac{1}{2}\dot{\phi}^{2}+V\left( \phi \right) +16\left( H^{2}+\dot{H%
}\right) Hf_{,\phi }\dot{\phi}+8H\left( \dot{\phi}^{2}f_{,\phi \phi
}+f_{,\phi }\ddot{\phi}\right) .
\end{equation}

In the following for the scalar field potential, we assume that $V\left( \phi
\right) =V_{0}e^{\lambda \phi }$, where in the case $\lambda =0$, $V\left(
\phi \right) $ describes the cosmological constant.

We employ dimensionless variables to rewrite the field equations in the
equivalent form of an algebraic-differential system of first-order
differential equations. To understand the evolution of the physical
variables and reconstruct the cosmological history provided by this
gravitational model, we investigate the existence of asymptotic solutions within the phase space.

Specifically, we calculate the stationary points of the field equations and
analyze their stability properties. Each stationary point corresponds to a
distinct epoch in cosmic evolution.

Now, consider the cosmological scenario where $f\left( \phi \right) $ is a
linear function, $f\left( \phi \right) =f_{0}\phi $, and $g\left( \phi
\right) =e^{2\beta \phi }$, an exponential function as described by the Chameleon mechanism or equivalently by the Weyl-integrable spacetime.

In this case, the components of the effective cosmological fluid are
expressed as \cite{dn4}%
\begin{equation}
\rho _{eff}=-\frac{1}{2}\dot{\phi}^{2}+V_{0}e^{\lambda \phi }-\rho
_{m0}e^{2\beta }a^{-3}\,+24f_{0}H\dot{\phi},
\end{equation}%
\begin{equation}
p_{eff}=\frac{1}{2}\dot{\phi}^{2}+V_{0}e^{\lambda \phi }+16f_{0}\left( H^{2}+%
\dot{H}\right) H\dot{\phi}+8Hf_{0}\ddot{\phi}.
\end{equation}%
and the equation of motion for the scalar field (\ref{ai.13}) becomes
\begin{equation}
0=\ddot{\phi}+3H\dot{\phi}-\lambda V_{0}e^{\lambda \phi }-24Hf_{0}\left(
H^{2}+\dot{H}\right) -2\beta e^{2\beta _{0}\phi }\rho _{m0}a^{-3}.
\end{equation}

\section{Dynamical System Analysis in the Hubble normalization}

\label{sec4}

We shall now consider the following dimensionless variables within the
Hubble normalization approach \cite{cop1,cop2}%
\begin{equation}
\eta ^{2}=\frac{H^{2}}{1+H^{2}}\ ,\ y^{2}=\frac{V\left( \phi \right) }{%
(1+H^{2})}\ ,\ x=\frac{\dot{\phi}}{\sqrt{(1+H^{2})}}\ ,\ \Omega _{m}=\frac{%
\rho _{m0}a^{-3}e^{2\beta \phi }}{1+H^{2}}
\end{equation}%
and 
\begin{equation}
\lambda =\frac{V^{\prime }\left( \phi \right) }{V\left( \phi \right) }\
,\Gamma (\lambda )=\frac{V\left( \phi \right) V^{\prime \prime }\left( \phi
\right) }{(V^{\prime }\left( \phi \right) )^{2}}\ 
\end{equation}

Using them and combining with equation (\ref{ai.11}) we obtain our
constraint equation, namely: 
\begin{equation}
\Omega _{m}=\frac{x^{2}}{2}-y^{2}+3\eta ^{2}-24f_{0}\frac{x\eta ^{3}}{1-\eta
^{2}}
\end{equation}%
For the exponential potential of the form $V=V_{0}e^{\lambda \phi }$, we
have that $\Gamma =1$ and $\lambda =const$. Thus, we end with the
three-dimensional autonomous system that describes the dynamics of this
model, which is the following

\begin{multline}
\frac{dx}{d\ln a}=\frac{1}{4 \left(-1 + 2 \eta^2 + \left(-1 + 96 f_0^2\right)
\eta^4 - 8 f_0 x \eta \left(-1 + \eta^2\right)\right)}\times \\
\Bigg( -2 y^2 \left(-1 + \eta^2\right) \Big( -2 \lambda + 2 \left(-12 f_0 +
\lambda\right) \eta^2 + x \eta \left(1 + 16 f_0 \lambda + \left(-1 + 8 f_0
\lambda\right) \eta^2\right)\Big) \\
+ \eta \Big( -48 f_0 \eta^3 \left(-1 + \eta^2\right) + x^3 \left(-1 +
\eta^2\right)^2 + 8 f_0 x^2 \eta \left(-15 + 13 \eta^2 + 2 \eta^4\right) \\
+ 6 x \left(2 - 5 \eta^2 + 4 \eta^4 + \left(-1 + 64 f_0^2\right)
\eta^6\right)\Big) \\
- 8 \beta \left(-1 + \eta^2\right) \left(-1 + \eta^2 + 4 f_0 x \eta \left(2
+ \eta^2\right)\right) \Omega_m \Bigg)
\end{multline}

\begin{multline}
\frac{dy}{d\ln a}=\frac{1}{2} y \Bigg( \lambda x + \Bigg( \eta \Big( 16 f_0 x
\eta^3 \left(-1 + \eta^2\right) + x^2 \left(-1 + \eta^2\right)^2 - 2 y^2
\left(-1 + \eta^2\right) \left(1 + \left(-1 + 8 f_0 \lambda\right)
\eta^2\right) \\
+ 2 \eta^2 \Big( -3 + 6 \eta^2 + 3 \left(-1 + 64 f_0^2\right) \eta^4 - 16
f_0 \beta \left(-1 + \eta^2\right) \Omega_m \Big)\Big)\Bigg) \\
\Big/ \Big( 2 \Big( -1 + 2 \eta^2 + \left(-1 + 96 f_0^2\right) \eta^4 - 8
f_0 x \eta \left(-1 + \eta^2\right) \Big) \Big) \Bigg)
\end{multline}

\begin{multline}
\frac{d \eta}{d \ln a}=\frac{ \left(-1 + \eta^2\right) \Big( + 2 \eta^2 \Big( %
-3 + 6 \eta^2 + 3 \left(-1 + 64 f_0^2\right) \eta^4 - 16 f_0 \beta \left(-1
+ \eta^2\right) \Omega_m \Big) \Big)}{ 4 \Big( -1 + 2 \eta^2 + \left(-1 + 96
f_0^2\right) \eta^4 - 8 f_0 x \eta \left(-1 + \eta^2\right) \Big)} \\
+ \frac{- 2 y^2 \left(-1 + \eta^2\right) \left(1 + \left(-1 + 8 f_0
\lambda\right) \eta^2\right)+16 f_0 x \eta^3 \left(-1 + \eta^2\right) + x^2
\left(-1 + \eta^2\right)^2 }{ 4 \Big( -1 + 2 \eta^2 + \left(-1 + 96
f_0^2\right) \eta^4 - 8 f_0 x \eta \left(-1 + \eta^2\right) \Big)}
\end{multline}

While the expression for the equation of state parameter can be written as 
\begin{multline}
w_{eff}=\frac{64f_{0}x\eta ^{3}\left( -1+\eta ^{2}\right) +x^{2}\left(
-1+\eta ^{2}\right) ^{2}-32f_{0}\eta ^{2}\Big(6f_{0}\eta ^{4}+\beta \left(
-1+\eta ^{2}\right) \Omega _{m}\Big)}{6\eta ^{2}\Big(-1+2\eta ^{2}+\left(
-1+96f_{0}^{2}\right) \eta ^{4}-8f_{0}x\eta \left( -1+\eta ^{2}\right) \Big)}
\\
-\frac{-2y^{2}\left( -1+\eta ^{2}\right) \left( 1+\left( -1+8f_{0}\lambda
\right) \eta ^{2}\right) }{6\eta ^{2}\Big(-1+2\eta ^{2}+\left(
-1+96f_{0}^{2}\right) \eta ^{4}-8f_{0}x\eta \left( -1+\eta ^{2}\right) \Big)}
\end{multline}

Proceeding with the dynamical analysis, we obtain the system's critical points corresponding to different cosmic eras. The results are presented
in Table \ref{tab:mexp1}. In the following lines, we discuss the physical properties of asymptotic solutions at 
stationary points.

\begin{table}[tbp] \centering%
\caption{Stationary points in the Hubble normalization.}%
\begin{tabular}{cccccc}
\hline\hline
\textbf{Point} & $\mathbf{\ }\left( \mathbf{x,y,\eta }\right) $ & \textbf{%
Existence} & $w_{eff}$ & \textbf{Acceleration} & \textbf{Attractor} \\ \hline
$P_{0}$ & $(0,0,0)$ & Always & Undefined & False & False \\ 
$P_{1}^{-}$ & $(0,0,-1)$ & Always & $-\frac{1}{3}$ & False & True \\ 
$P_{1}^{+}$ & $(0,0,1)$ & Always & $-\frac{1}{3}$ & False & False \\ 
$P_{2}^{-}$ & $\left(-\frac{1}{6\beta },0,-1\right)$ & $\beta \neq 0$ & $-\frac{5}{9}$
& True & False \\ 
$P_{2}^{+}$ & $\left(\frac{1}{6\beta },0,1\right)$ & $\beta \neq 0$ & $-\frac{5}{9}$ & 
True & False \\ 
$P_{3}^{+}$ & $\left(3\sqrt{\frac{2}{15-20\sqrt{30}f_{0}}},0,-\sqrt{\frac{3}{3-4%
\sqrt{30}f_{0}}}\right)$ & $f_{0}<\frac{\sqrt{\frac{3}{10}}}{4}$ & $-1$ & de Sitter
& False \\ 
$P_{3}^{-}$ & $\left(-3\sqrt{\frac{2}{15-20\sqrt{30}f_{0}}},0,\sqrt{\frac{3}{3-4%
\sqrt{30}f_{0}}}\right)$ & $f_{0}<\frac{\sqrt{\frac{3}{10}}}{4}$ & $-1$ & de Sitter
& True in regions \\ 
$P_{4}^{-}$ & $\left(-3\sqrt{\frac{2}{15+20\sqrt{30}f_{0}}},0,\sqrt{\frac{3}{3+4%
\sqrt{30}f_{0}}}\right)$ & $f_{0}>-\frac{\sqrt{\frac{3}{10}}}{4}$ & $-1$ & de
Sitter & False \\ 
$P_{4}^{+}$ & $\left(-3\sqrt{\frac{2}{15+20\sqrt{30}f_{0}}},0,\sqrt{\frac{3}{3+4%
\sqrt{30}f_{0}}}\right)$ & $f_{0}>-\frac{\sqrt{\frac{3}{10}}}{4}$ & $-1$ & de
Sitter & True in regions \\ 
$P_{5}^{-}$ & $\left(-\frac{3\sqrt{-3+8\beta ^{2}}}{2\beta \sqrt{-3+8\beta
(8f_{0}+\beta )}},0,-\frac{\sqrt{-3+8\beta ^{2}}}{\sqrt{-3+8\beta
(8f_{0}+\beta )}}\right)$ & some regions & $-1$ & de Sitter & False \\ 
$P_{5}^{+}$ & $\left(\frac{3\sqrt{-3+8\beta ^{2}}}{2\beta \sqrt{-3+8\beta
(8f_{0}+\beta )}},0,\frac{\sqrt{-3+8\beta ^{2}}}{\sqrt{-3+8\beta
(8f_{0}+\beta )}}\right)$ & some regions & $-1$ & de Sitter & True in regions \\ 
$P_{6}^{-}$ & $\left(0,\sqrt{\frac{3\lambda }{-8f_{0}+\lambda }},-\sqrt{\frac{%
\lambda }{-8f_{0}+\lambda }}\right)$ & some regions & $-1$ & de Sitter & False
\\ 
$P_{6}^{+}$ & $(0,\sqrt{\frac{3\lambda }{-8f_{0}+\lambda }},\sqrt{\frac{%
\lambda }{-8f_{0}+\lambda }})$ & some regions & $-1$ & de Sitter & True in
regions \\ \hline\hline
\end{tabular}%
\label{tab:mexp1}%
\end{table}%

The point $P_{0}~$ corresponds to an empty universe, the Minkowski space, and the
point describes an unstable solution.

$P_{1}^{-}$ describes a universe that is dominated by the Gauss-Bonnet
scalar, the asymptotic solution is a stable scaling solution with $w_{eff}=-%
\frac{1}{3}$ and eigenvalues $\left( -2,-1,-1\right) $.

On the other hand, point $P_{1}^{-}$ with $w_{eff}=-\frac{1}{3}$ describes
the same physical solution as $P_{1}^{-}$, but the point is unstable since its eigenvalues are ( 1, 1, 2 ).

Stationary point $P_{2}^{-}$ describes a universe where interaction exists,
the effective equation of state parameter is $w_{eff}=-\frac{5}{9}$ which
describes an accelerating universe and its  eigenvalues are $\left( -\frac{4}{3},1,-\frac{%
8\beta +\lambda }{12\beta }\right) $ from where we infer that the point is
always saddle.

Similarly, point $P_{2}^{+}$ has the same physical properties with $%
P_{2}^{-} $, the eigenvalues are $\left( \frac{8}{3},3,\frac{16\beta
-\lambda }{12\beta }\right) $, from where infer that $P_{2}^{+}$ is a source
for $\frac{16\beta -\lambda }{12\beta }>0$, otherwise is a saddle point.

The existence condition for point $P_{3}^{+}$, is $f_{0}<\frac{1}{4}\sqrt{%
\frac{3}{10}}\,$, the asymptotic solution describes the de Sitter universe, that is, $w_{eff}=-1$. The eigenvalues are calculated to be $(\frac{3\sqrt{-\frac{1%
}{A}}(B-C)}{5f_{0}A},\frac{3\sqrt{-\frac{1}{A}}(B+C)}{5f_{0}A},\frac{3\sqrt{-%
\frac{1}{A}}\lambda }{\sqrt{10}})$, where $A=-3+4\sqrt{30}f_{0},B=-15\sqrt{3}%
f_{0}+60\sqrt{10}f_{0}^{2}-3\sqrt{10}f_{0}\beta +40\sqrt{3}f_{0}^{2}\beta ,C=%
\sqrt{90f_{0}^{2}\beta ^{2}-240\sqrt{30}f_{0}^{3}\beta
^{2}+4800f_{0}^{4}\beta ^{2}}$. Hence, the stationary point is an attractor for $B-C>0$,~$B+C>0$ and $\lambda <0$.\ It follows that $P_{3}^{+}$ is a source for $\left\{ \beta >-\frac{1}{2}\sqrt{\frac{15}{2}},\lambda >0:f_{0}<%
\frac{1}{4}\sqrt{\frac{3}{10}},f_{0}\neq 0\right\} $, otherwise it is a saddle point.

Furthermore, point $P_{3}^{-}$ has the same existence condition and
effective equation of state parameter with point $P_{3}^{+};$ however, the
eigenvalues are $(\frac{-3\sqrt{-\frac{1}{A}}(B+C)}{5f_{0}A},\frac{3\sqrt{-%
\frac{1}{A}}(-B+C)}{5f_{0}A},-\frac{3\sqrt{-\frac{1}{A}}\lambda }{\sqrt{10}}%
) $ from where we infer that the point is an attractor for $B-C<0$,~$B+C<0$
and $\lambda >0$, that is, $P_{3}^{-}$ is an attractor for the region where $%
P_{3}^{+}$ is a source.

Point $P_{4}^{+}$ it is real for $f_{0}>-\frac{\sqrt{\frac{3}{10}}}{4}$, and describes a de Sitter universe, that is, $w_{eff}=-1$, the stability
condition gives that the point is stable for $\left\{ \beta <\frac{1}{2}%
\sqrt{\frac{15}{2}},\lambda <0:\left\vert f_{0}\right\vert <\frac{1}{4}\sqrt{%
\frac{3}{10}},f_{0}\neq 0\right\} .$

On the other hand, the stationary point $P_{4}^{-}$ always describes an
unstable de Sitter solution.

The stationary points $P_{5}^{+}$ and $P_{5}^{-}$ describes de Sitter
solutions with nonzero interacting term. The points are real when the free
parameters are constraint as $\left( \ f_{0}<\frac{3-8\beta ^{2}}{64\beta }%
,~\beta <-\sqrt{\frac{3}{8}},~:0<\beta <\sqrt{\frac{3}{8}}\right) $ or$%
\left( \ f_{0}>\frac{3-8\beta ^{2}}{64\beta },~-\sqrt{\frac{3}{8}}<\beta
<0,~:\sqrt{\frac{3}{8}}<\beta \right) $\thinspace . Stationary point $%
P_{5}^{-}$ describes always an unstable solution, however, $P_{5}^{+}$ is an
attractor when 
\begin{equation*}
\left( f_{0}<0,\lambda >0:-\beta _{1}<\beta <-\sqrt{\frac{15}{8}},~-\beta
_{2}<\beta <-\sqrt{\frac{3}{8}}\right) ,
\end{equation*}%
\begin{equation*}
\left( 0<f_{0}<\frac{3-8\beta ^{2}}{64\beta },\lambda >0:-\beta _{1}<\beta <-%
\sqrt{\frac{15}{8}},~-\beta _{2}<\beta <-\sqrt{\frac{3}{8}}\right) ,
\end{equation*}%
\begin{equation*}
\left( f_{0}>0,\lambda <0:\sqrt{\frac{15}{8}}<\beta <\beta _{1},~\sqrt{\frac{%
3}{8}}<\beta <\beta _{2}\right) ,
\end{equation*}%
where $\beta _{1}=\frac{\sqrt{42\left( 43+4\sqrt{74}\right) }}{28}$ and $%
\beta _{2}=\frac{\sqrt{42\left( 43-4\sqrt{74}\right) }}{28}$.  

Finally, the two stationary points $P_{6}^{+}$ and $P_{6}^{-}$ are real when 
$\lambda f_{0}<0$. These stationary points describe de Sitter solutions. The
eigenvalue analysis reveals that $P_{6}^{+}$ is an attractor for $f_{0}<0$
and $0<\lambda <\sqrt{\frac{6}{17}}$, while under the same conditions, point 
$P_{6}^{-}$ acts as a source. Otherwise, these stationary points behave as
saddle points.

In the Hubble normalization approach, the dynamical variables are not
compactified, meaning that they can also take values in the infinite regime. The analysis we presented considers only the finite regime of the dynamical variables. It is necessary to employ compactified variables to understand the evolution and behavior of the solutions at infinity. Instead of introducing compactified variables within the Hubble normalization approach, we continue our study by defining new dimensionless variables that are compactified.

\begin{figure}[tbp]
\centering
\includegraphics[width=1\linewidth]{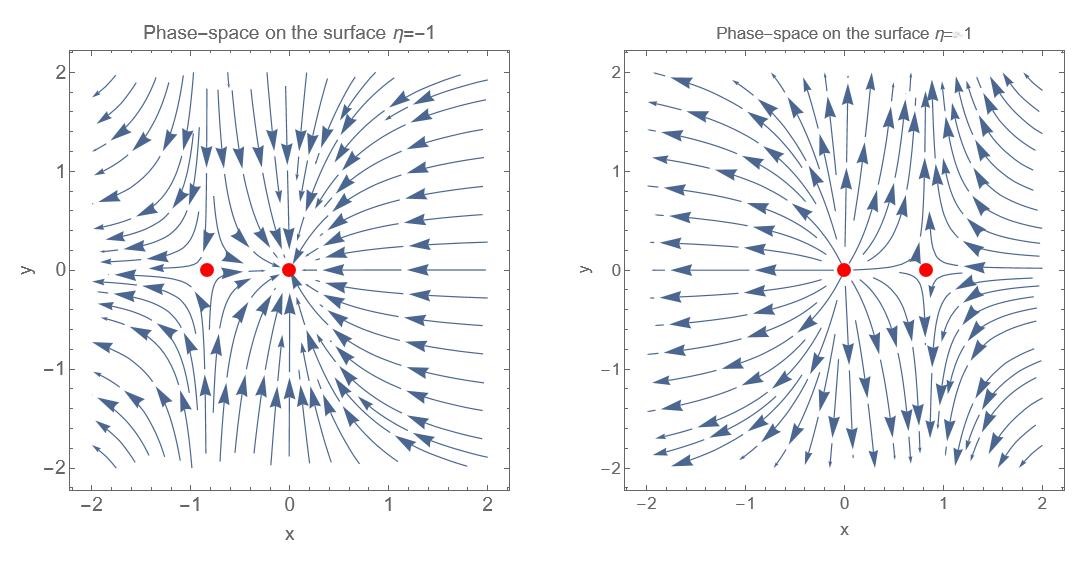}
\caption{"x", "y" Phase-space on the surface $\protect\eta=-1$, assuming $%
f_{0} \rightarrow 10, \protect\beta \rightarrow \frac{1}{5}, \protect\lambda %
\rightarrow 1$}
\label{fig:enter-label}
\end{figure}
\begin{figure}[tbp]
\centering
\includegraphics[width=0.5\linewidth]{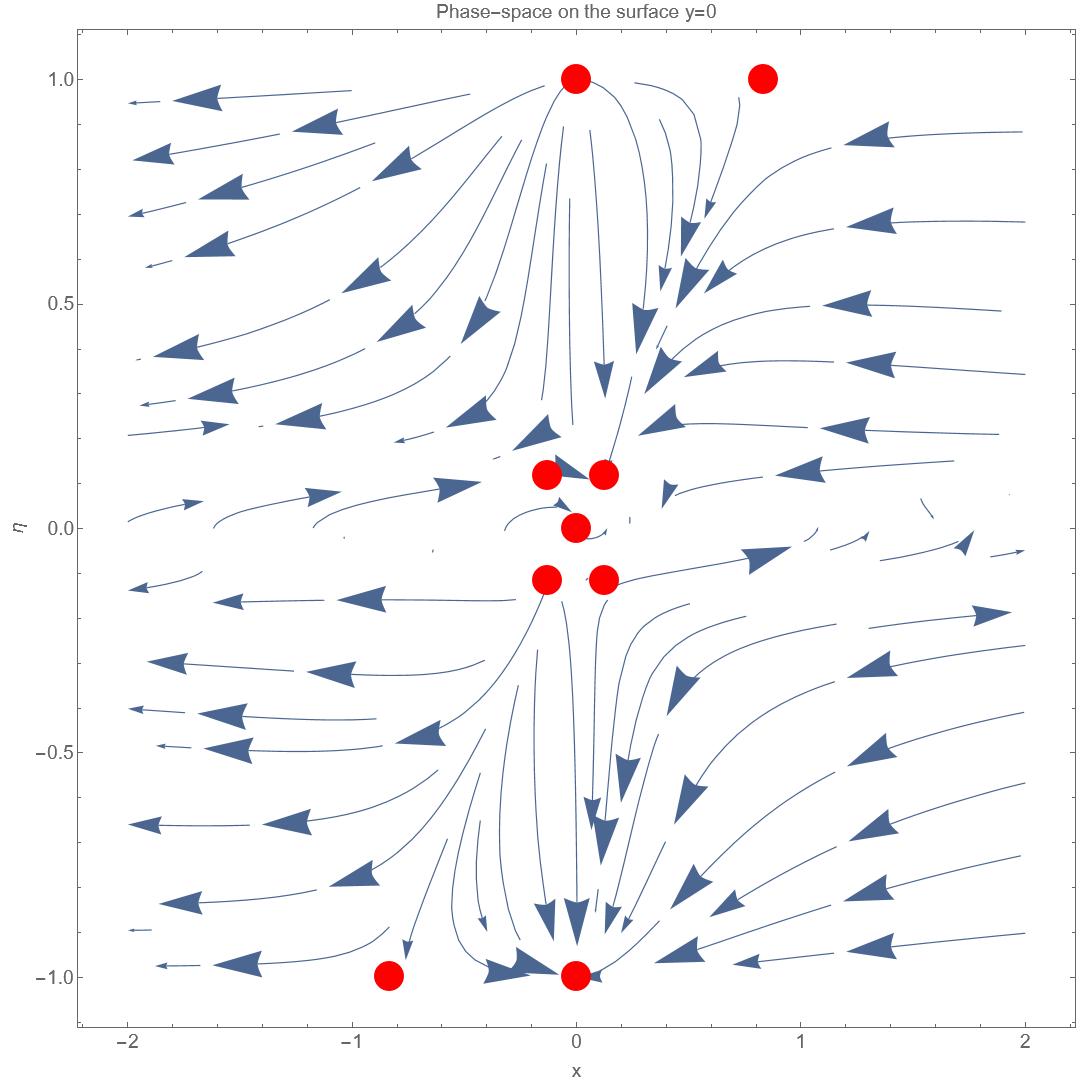}
\caption{"x", "$\protect\eta$" Phase-space on the surface $y=0$, assuming $%
f_{0} \rightarrow -10, \protect\beta \rightarrow \frac{1}{5}, \protect%
\lambda \rightarrow 1$}
\label{fig:enter-label}
\end{figure}

\section{Dynamical System Analysis in matter-scalar field normalization}

\label{sec5}

We consider a new set of dimensionless variables which were introduced
recently in \cite{angenl}, for analyzing phantom-scalar field dynamics.

Specifically, we introduce the dimensionless variables:%
\begin{equation}
\chi =\frac{\dot{\phi}}{D},~\zeta ^{2}=\frac{V\left( \phi \right) }{D^{2}}%
,~\omega _{m}=\frac{\rho _{m}e^{2\beta \phi }}{D^{2}},~r=\frac{H}{D},~\Phi =%
\frac{f_{0}}{D^{2}}
\end{equation}%
where function $D$ is defined as 
\begin{equation}
D=\sqrt{\frac{1}{2}\dot{\phi}^{2}+V\left( \phi \right) +\rho _{m}e^{2\beta
\phi }}.
\end{equation}

In terms of these variables, the constraint equation reads%
\begin{equation}
\omega _{m}=3r^{2}\left( 1-8rx\Phi \right) +\frac{1}{2}\chi ^{2}-\zeta ^{2}.
\end{equation}%
Variables $\chi $ and $\zeta $ are compactified and constrained by $-\sqrt{2}%
\leq \chi \leq \sqrt{2}$, and $0\leq \zeta \leq 1$.

Moreover, from the definition of the dimensionless variables, the constraint equation follows
\begin{equation}
\frac{1}{2}\chi ^{2}+\zeta ^{2}+\omega _{m}=1.
\end{equation}

By definition, $D\neq 0$, meaning there will always be a matter contribution in the solution. Consequently, this normalization does not recover solutions without contribution from the matter term in the cosmological fluid. On the other hand, it enables the determination of stationary points
that describe scaling solutions, where the kinetic term of the scalar field
contributes to the cosmological fluid.

We introduce the new independent variable $dt=Dd\sigma $, and the field
equations reads%
\begin{align}
\frac{d\chi }{d\sigma } &=\chi \frac{d}{d\sigma }\left( \frac{1}{D}\right) +%
\frac{-\lambda \zeta ^{2}+12r^{4}\Phi -6r^{2}\left( 5\chi ^{2}+2\zeta
^{2}\right) \Phi -2\beta \omega _{m}+r\chi \left( 3+8\Phi \left( \lambda
\zeta ^{2}+2\beta \omega _{m}\right) \right) }{8r\Phi \left( \chi
+12r^{3}\Phi \right) -1}, \\
\frac{d\zeta }{d\sigma } &=\zeta \frac{d}{d\sigma }\left( \frac{1}{D}%
\right) +\frac{\lambda }{2}\chi \zeta , \\
\frac{dr}{d\sigma } &=r\frac{d}{d\sigma }\left( \frac{1}{D}\right) -\frac{1%
}{4}\frac{\chi ^{2}+2\zeta ^{2}+2r^{2}\left( 8\Phi \left( \lambda \zeta
^{2}-\chi r+24r^{4}\Phi +2\beta \omega _{m}\right) -3\right) }{8r\Phi \left(
\chi +12r^{3}\Phi \right) -1}, \\
\frac{d\Phi }{d\sigma } &=-2\Phi \frac{d}{d\sigma }\left( \frac{1}{D}%
\right),
\end{align}%
where 
\begin{eqnarray}
\frac{2}{D}\frac{dD}{d\sigma } &=&\lambda \chi \zeta ^{2}-\left( 3r-2\beta
\chi \right) \omega _{m} \\
&&+\frac{\chi \left( -\lambda \zeta ^{2}+12r^{4}\Phi -6r^{2}\left( 5\chi
^{2}+2\zeta ^{2}\right) \Phi -2\beta \omega _{m}+r\chi \left( 3+8\Phi \left(
\lambda \zeta ^{2}+2\beta \omega _{m}\right) \right) \right) }{8r\Phi \left(
\chi +12r^{3}\Phi \right) -1}.
\end{eqnarray}

Moreover, the equation of state parameter reads%
\begin{equation*}
w_{eff}=\frac{\chi ^{2}+2\zeta ^{2}-64r^{3}\chi \Phi +16r^{2}\Phi \left(
\lambda \zeta ^{2}-12r^{4}\Phi +2\beta \omega _{m}\right) }{6r^{2}\left(
8r\Phi \left( \chi +12r^{3}\Phi \right) -1\right) }\text{.}
\end{equation*}

The stationary points of the latter dynamical system are presented in Table 

We observe that all the stationary points correspond to de Sitter asymptotic
solutions. Stationary points $Q_{1}^{\pm }$ and $Q_{2}^{\pm }$ are analogous
to points $P_{3}^{\pm }$ and $P_{4}^{\pm }$ in the Hubble normalization
approach. Similarly, points $Q_{4}^{\pm }$, where only the scalar field
potential contributes to the cosmic fluid, correspond to points $P_{6}^{\pm
} $. Furthermore, points $Q_{3}^{\pm }$, characterized by non-zero
interaction, describe asymptotic solutions with the same physical properties
as points $P_{5}^{\pm }$. Due to this analogy, we omit the detailed
presentation of the stability properties for these points.

The most significant result of this normalization is that the stationary
points at the finite regime, that is, points where $\chi ^{2}\rightarrow 
\sqrt{2}$, $Q_{1}^{\pm }$ and $Q_{2}^{\pm }$ and points with $\zeta
\rightarrow 1$, $Q_{4}^{\pm }~$describe de Sitter solutions. There are
no Big Rip or Big Crunch singularities. In other words, there are no
stationary points with $w_{eff}= \pm \infty$. We conclude that the
inclusion of the Gauss-Bonnet term in the gravitational field equations
prevents the occurrence of future or past singularities. This conclusion
also holds without interaction between the scalar field and the
matter source, i.e., for $\beta =0$.

This result contrasts with General Relativity, where in the absence of
interaction, Big Rip singularities serve as future attractors for the
phantom field \cite{q15}. When the scalar field is present, Big Rip
singularities remain as stationary points but describe unstable solutions 
\cite{q14b}.

\begin{table}[tbp] \centering%
\caption{Stationary points in the matter scalar -field normalization.}%
\begin{tabular}{ccccc}
\hline\hline
\textbf{Point} & $\mathbf{\ }\left( \mathbf{\omega }_{m},\mathbf{\chi ,\zeta
,r,\Phi }\right) $ & \textbf{Existence} & $w_{eff}$ & \textbf{Acceleration}
\\ 
$Q_{1}^{-}$ & $\left( 0,\sqrt{2},0,-\sqrt{\frac{5}{3}},-\frac{3}{20}\sqrt{%
\frac{3}{10}}\right) $ & Always & $-1$ & de Sitter \\ 
$Q_{1}^{+}$ & $\left( 0,\sqrt{2},0,\sqrt{\frac{5}{3}},-\frac{3}{20}\sqrt{%
\frac{3}{10}}\right) $ & Always & $-1$ & de Sitter \\ 
$Q_{2}^{-}$ & $\left( 0,-\sqrt{2},0,-\sqrt{\frac{5}{3}},\frac{3}{20}\sqrt{%
\frac{3}{10}}\right) $ & Always & $-1$ & de Sitter \\ 
$Q_{2}^{+}$ & $\left( 0,-\sqrt{2},0,\sqrt{\frac{5}{3}},-\frac{3}{20}\sqrt{%
\frac{3}{10}}\right) $ & Always & $-1$ & de Sitter \\ 
$Q_{3}^{-}$ & $\left( 1+\frac{6}{8\beta ^{2}-21},-\frac{2}{3\sqrt{8\beta
^{2}-21}},0,-\frac{4\beta }{\sqrt{3\left( 8\beta ^{2}-21\right) }},-\frac{%
3\left( 63+64\beta ^{2}\left( \beta ^{2}-3\right) \right) }{1024\beta ^{2}}%
\right) $ & Always & $-1$ & de Sitter \\ 
$Q_{3}^{+}$ & $\left( 1+\frac{6}{8\beta ^{2}-21},\frac{2}{3\sqrt{8\beta
^{2}-21}},0,\frac{4\beta }{\sqrt{3\left( 8\beta ^{2}-21\right) }},-\frac{%
3\left( 63+64\beta ^{2}\left( \beta ^{2}-3\right) \right) }{1024\beta ^{2}}%
\right) $ & Always & $-1$ & de Sitter \\ 
$Q_{4}^{-}$ & $\left( 0,0,1,-\frac{1}{\sqrt{3}},-\frac{3}{8}\lambda \right) $
& Always & $-1$ & de Sitter \\ 
$Q_{4}^{+}$ & $\left( 0,0,1,\frac{1}{\sqrt{3}},-\frac{3}{8}\lambda \right) $
& Always & $-1$ & de Sitter \\ \hline\hline
\end{tabular}%
\label{tabl2}%
\end{table}%

\section{Conclusions}

\label{sec6}

Within the four-dimensional Einstein-Gauss-Bonnet scalar field, spatially
flat FLRW cosmology, we consider a phantom scalar field with a negative
kinetic energy and a dust fluid that exhibits nonzero energy transfer with
the scalar field. To understand how the Gauss-Bonnet scalar modifies the
behavior and evolution of the physical parameters, we perform a detailed
phase-space analysis using two different sets of dimensionless variables.
The introduction of these two sets of variables is essential for conducting a
comprehensive phase-space analysis and to explore the existence of
stationary points in the extreme limits.

The first normalization belongs to the family of Hubble normalizations.
Within these variables, the stationary points of the cosmological field
equations describe the vacuum Minkowski solution, scaling solutions where
the Gauss-Bonnet term contributes to the cosmological dynamics, and de Sitter solutions that can act as future attractors. Conversely, the second
set of dimensionless variables is based on a matter-scalar field
normalization. This set has been introduced to investigate the evolution of
the physical properties in the extreme limits of the scalar field
components. The stationary points this normalization supports include
those with nonzero contributions from both matter and the scalar field.
Consequently, in the extreme limit, only de Sitter solutions are possible,
indicating the absence of asymptotic solutions describing Big Rip or Big
Crunch singularities.

Although we consider a nonzero interaction between the scalar field and the
dust fluid, the main conclusions of this work remain valid, even in the
absence of the interaction term. Specifically, in the limit where the
parameter $\beta \to 0$, only the stationary points $P_{2}^{\pm}$ cease to
exist. Moreover, for the three free functions of the gravitational model, we
consider specific forms that reduce the dimensionality of the field
equations. However, our main conclusions hold for other functional
forms of the free functions, namely the two coupling functions $f\left( \phi
\right)$, $g\left( \phi \right)$, and the scalar field potential $V\left(
\phi \right)$.

\begin{acknowledgments}
AP\ \&\ GL thanks the support of VRIDT through Resoluci\'{o}n VRIDT No.
096/2022 and Resoluci\'{o}n VRIDT No. 098/2022. This study was supported by
FONDECYT 1240514, Etapa 2024.
\end{acknowledgments}


\begin{thebibliography}{99}
\bibitem{pan1} T. Padmanabhan and D. Kothawala, Lanczos-Lovelock models of
gravity, Phys. Reports 531, 115 (2013)

\bibitem{rr1} P.G.S. Fernandes, P. Carrilho, T. Clifton and D.J. Mulryne,
Class. Quantum Grav. 39, 063001 (2022)

\bibitem{lov1} D. Lovelock, J. Math. Phys. 12, 498 (1971)

\bibitem{ost1} M. Crisostomi, R. Klein, and D. Roest, J. High Energy Phys.
06, 124 (2017)

\bibitem{gb1} D. Chirkov, S. A. Pavluchenko and A. Toporensky, Gen.\ Rel.
Grav. 46, 1799 (2014)

\bibitem{gb2} C. Charmousis, J.-F. Dufaux, Class. Quantum Grav. 19, 4671
(2002)

\bibitem{gb3} J. Novak, M. Ozkan, Y. Pang, G. Tartaglino-Mazzucchelli,
Phys.\ Rev. Lett. 119, 111602 (2017)

\bibitem{gb4} E.O. Pozdeeva, M.A. Skugoreva, A.V. Toporensky and S. Yu
Vernov, JCAP 09, 050 (2024)

\bibitem{gb5} G. Dotti, J. Oliva and R. Troncoso, Phys. Rev. D 76, 064038
(2007)

\bibitem{gb6} A. Giacomini, J. Oliva and A. Vera, Phys. Rev. D 91, 104033
(2015)

\bibitem{gb7} S. Hansraj, N. Mkhize, Phys.\ Rev. D\ 102, 084028 (2020)

\bibitem{gb8} G. Kofinas, Class. Quantum Grav. 22, L47 (2005)

\bibitem{gbs1} S. Tsujikawa and M. Sami, JCAP 0701, 006 (2007)

\bibitem{gbs2} T. Koivisto and D.F. Mota, Phys. Lett. B 644, 104 (2007)

\bibitem{gbs3} M. Sami, A. Toporensky, P. V. Tretjakov and S. Tsujikawa,
Phys. Lett. B 619, 193 (2005)

\bibitem{gbb1} I.V. Fomin, Phys. Part. Nucl. 49, 525 (2018)

\bibitem{gbb2} I.V. Fomin and S.V.\ Chernov, Grav. Cosmo. 23, 367 (2017)

\bibitem{gbb3} S.D. Odintsov and V.K. Oikonomou, Phys. Rev.\ D 98, 044039
(2018)

\bibitem{gbb4} I. Supriyadi and G. Hikmawan, J. Phys. Conf. Ser. 2243,
012095 (2022)

\bibitem{gbb5} S. Chakraborty, T. Paul and S. SenGupta, Phys.\ Rev. D 98,
083539 (2018)

\bibitem{kant3} M. Motaharfar and H.R. Sepangi, Eur. Phys. J. C 76, 646
(2016)

\bibitem{kant4} N. Rashidi and K. Nozari, The Astrophysical Journal, 890, 58
(2020)

\bibitem{in4} S.D. Odintsov, V.K. Oikonomou and F.P. Fronimos, Annals of
Physics 420, 168250 (2020)

\bibitem{dn3} A.D. Milano, G. Leon and A. Paliathanasis, Mathematics 11,
1408 (2023)

\bibitem{dn4} A.D. Milano, G. Leon and A.\ Paliathanasis, Phys. Rev. D 108,
023519 (2023)

\bibitem{dn6} V.K. Oikonomou, P.\ Tsyba and O. Razina, Annals Phys. 462,
169597 (2024)

\bibitem{dn7} M.M. Gohain, K. Bhuyan, Phys.\ Scr. 99, 075306 (2024)

\bibitem{dn8} B. Micolta-Riascos, A.D. Milano, G.\ Leon, B. Droguett, E.
Gonzalez and J. Magana, Fractal Fract. 8, 626 (2024)

\bibitem{ch1} J. Khoury and A. Wetlman,Phys. Rev. Lett. 93, 171104 (2004)

\bibitem{ch2} J. Khoury and A. Wetlman, Phys.\ Rev. D 69, 044026 (2004)

\bibitem{q14} R.C. Caldwell, M. Kamionkowski and N.N. Weinberg, Phys. Rev.
Lett. 91, 071301 (2003)

\bibitem{qq1} L.P. Chimento and R. Lazkoz, Phys.\ Rev. Lett. 91, 211301
(2003)

\bibitem{qq2} L.A. Ure\~{n}a-L\'{o}pez, JCAP 09, 013 (2005)

\bibitem{qq3} N. Roy, S. Goswami and S. Das, Phys.\ Dark\ Univ. 36, 101037
(2022)

\bibitem{q14a} R. Curbelo, T. Gonzalez, G. Leon and I. Quiros, Class.
Quantum Grav. 23, 1585 (2006)

\bibitem{q14b} A. Paliathanasis, A. Halder and G. Leon, arXiv:2412.06501
(2024)

\bibitem{q15} V. Faraoni, Phantom cosmology with general potentials, Class.
Quantum Grav. 22, 3235 (2005)

\bibitem{anph} A. Paliathanasis, Gen. Rel. Grav. 56, 84 (2024)

\bibitem{dn22} A.A. Coley, Dynamical Systems in Cosmology (1999)
[gr-qc/9910074]

\bibitem{dn33} G.\ Leon, Class.\ Quantum Grav. 26, 035008 (2009)

\bibitem{dn44} S. Chatzidakis, A. Giacomini, P.G.L. Leach, G. Leon and A.
Paliathanasis, J. High Energ. Astroph. 36, 141 (2022)

\bibitem{dn55} R. De Arcia, I. Quiros, U. Nucamendi and T. Gonzales, Phys.\
Dark Univ. 40, 101183 (2023)

\bibitem{cop1} E.J. Copeland, A.R. Liddle and D.\ Wands, Phys. Rev. D 57,
4686 (1998)

\bibitem{cop2} T. Gonzalez, G. Leon and I. Quiros, Class. Quantum Grav. 23,
3165 (2006)

\bibitem{angenl} G. Leon, D. Shankar, A. Halder and A. Paliathanasis,
arXiv:2501.09177 (2025)
\end{thebibliography}
\end{document}